\documentclass[%
12pt,
reprint,
amsmath,amssymb,
floatfix,
superscriptaddress,
longbibliography
]{revtex4-1}
\usepackage{graphicx}% Include figure files
\usepackage{dcolumn}% Align table columns on decimal point
\usepackage{bm}% bold math
\usepackage{placeins} %FloatBarrier
\usepackage{miller}
\usepackage{siunitx} % 90\si{\degree}
\usepackage{tikz}
\usepackage{comment}
\usepackage{graphicx}
\usepackage{subfigure}

\usepackage[yyyymmdd,hhmmss]{datetime}

\usepackage[unicode=true, 
 bookmarks=true,bookmarksnumbered=false,bookmarksopen=false,
 breaklinks=true,pdfborder={0 0 1},backref=false,colorlinks=true]
 {hyperref}
\hypersetup{pdftitle={Diffractive triangulation of radiative point sources},
 pdfauthor={Stefano Vespucci, Aimo Winkelmann}}

\begin{document}

\title{Diffractive triangulation of radiative point sources}

\author{Stefano Vespucci}
\email{stefano.vespucci@strath.ac.uk}
\affiliation{Department of Physics, SUPA, University of Strathclyde, Glasgow G4 0NG, United Kingdom }

\author{Carol  Trager-Cowan}
%\email{stefano.vespucci@strath.ac.uk}
\affiliation{Department of Physics, SUPA, University of Strathclyde, Glasgow G4 0NG, United Kingdom }

\author{Dzmitry Maneuski}
\affiliation{School of Physics and Astronomy, SUPA, University of Glasgow, Glasgow G12 8QQ, United Kingdom }

\author{Val O'Shea}
\affiliation{School of Physics and Astronomy, SUPA, University of Glasgow, Glasgow G12 8QQ, United Kingdom }

\author{Aimo Winkelmann}
\email{aimo.winkelmann@bruker.com}
\affiliation{Bruker Nano GmbH, Am Studio 2D, 12489 Berlin, Germany }

\date{\today \, \currenttime}% It is always \today, today,

\begin{abstract}
We describe a general method to determine the location of a point source of waves relative to a two-dimensional active pixel detector.
Based on the inherent structural sensitivity of crystalline sensor materials, characteristic detector diffraction patterns can be used to triangulate the location of a wave emitter. 
As a practical application of the wide-ranging principle, a digital hybrid pixel detector is used to localize a source of electrons for Kikuchi diffraction pattern measurements in the scanning electron microscope. 
This provides a method to calibrate Kikuchi diffraction patterns for accurate measurements of microstructural crystal orientations, strains, and phase distributions. 
%\texttt{Version: \today\  \currenttime}
\end{abstract}

\maketitle

\begin{figure}[hbt!]   %
	\centering
	\includegraphics[width=0.77\columnwidth]{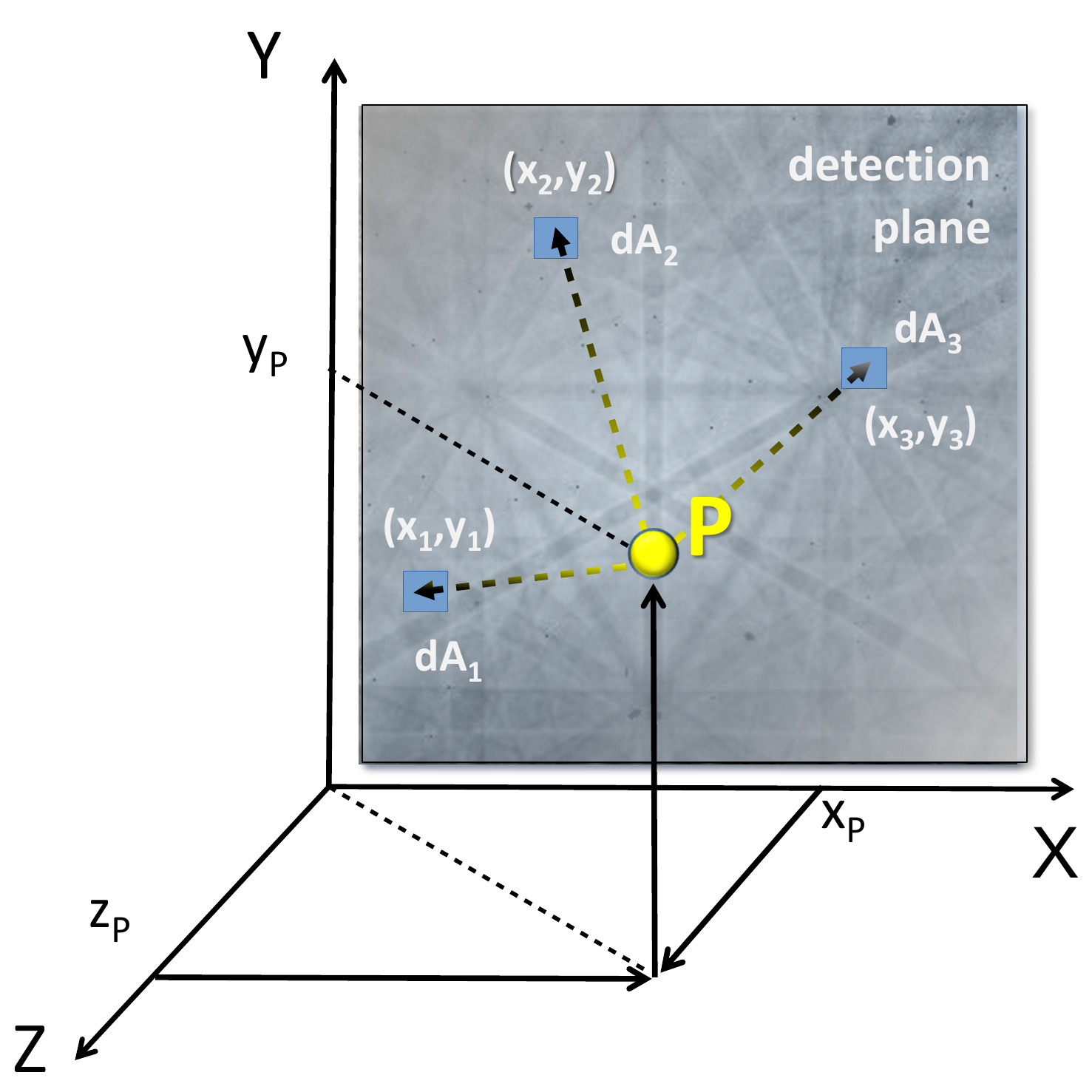}
	\caption{Principle of determination of the source coordinates 
		$(x_{\mathrm{P}},y_{\mathrm{P}},z_{\mathrm{P}})$.
		The detector with area elements $dA_n$ is reacting to diffraction effects of the waves from the source in the area element. 
		Each area element at a specific position $(x_n,y_n)$ on the detector is sensitive to the \textit{direction} from the source to the area element. 
		A triangulation procedure involving known diffraction features formed in the detector plane allows the source position to be determined.}
	\label{fig:principle}
\end{figure}

The accurate determination of the three-dimensional position of objects is connected to many measurement problems in the experimental sciences and in technological applications \cite{faugeras1993}. 
Very often, however, the object of interest is not directly accessible. 
In such situations, we can still obtain directional measurements from known reference points and then triangulate the position of the object. 
This trivial principle is illustrated in Fig.\,\ref{fig:principle}, where measurements of the angles from the two reference points $(x_1,y_1)$ and $(x_2,y_2)$  to the point \textbf{P} would be sufficient to determine the three-dimensional coordinates $(x_{\mathrm{P}},y_{\mathrm{P}},z_{\mathrm{P}})$ of that point, given that we know the reference distances in the $XY$-plane.

In this paper we will discuss an alternative concept of position determination, which specifically can be used to determine the position of localized sources of radiation  used in crystallography (e.g. electrons, X-rays, neutrons). 
In our method, instead of performing direct angular measurements from isolated reference points, an extended two-dimensional detector area is assumed to have a sensitivity which depends in a specific way on the incident direction of the waves on each area element $d\textbf{A}$ on the detector surface.
This angular sensitivity will be encoded by the internal crystalline structure of each area element, which can react to the specific \textit{wave-like} properties of the incident radiation.  
As a result of diffraction effects inside each pixel, the detector displays for each pixel area element $d\textbf{A}$ an intensity related to the direction from the area element $d\textbf{A}$ to the source point.
Each possible three-dimensional position of the source \textbf{P} relative to the detector defines a characteristic two-dimensional intensity pattern of the detector area elements. 
In contrast to direct angular triangulation measurements, the individual measurement points by themselves do not carry sufficient information to reconstruct the position of \textbf{P}.
Instead, in the method discussed in this paper, \textbf{P} is determined by the combined 2D signal of all detector pixels $d\textbf{A}_n$  which is illustrated by the characteristic pattern seen in the $XY$-plane in \mbox{Fig.\,\ref{fig:principle}}. 
With respect to the use of the information distributed in a two-dimensional pattern, this method can be seen as related to point source holography \cite{gabor1948nat,jericho2010}. Instead of imaging a 3D object as an interference pattern on a 2D detector, however, we wish to find the position of a localized incoherent source when the 2D pattern resulting from the diffraction in a known object (the detector) is given.
A calibration procedure then relates the three-dimensional position of the source at \textbf{P} and the corresponding projective two-dimensional features formed by the area elements $d\textbf{A}_n$. 
\begin{figure}   %
	\centering
	\includegraphics[width=1.0\columnwidth]{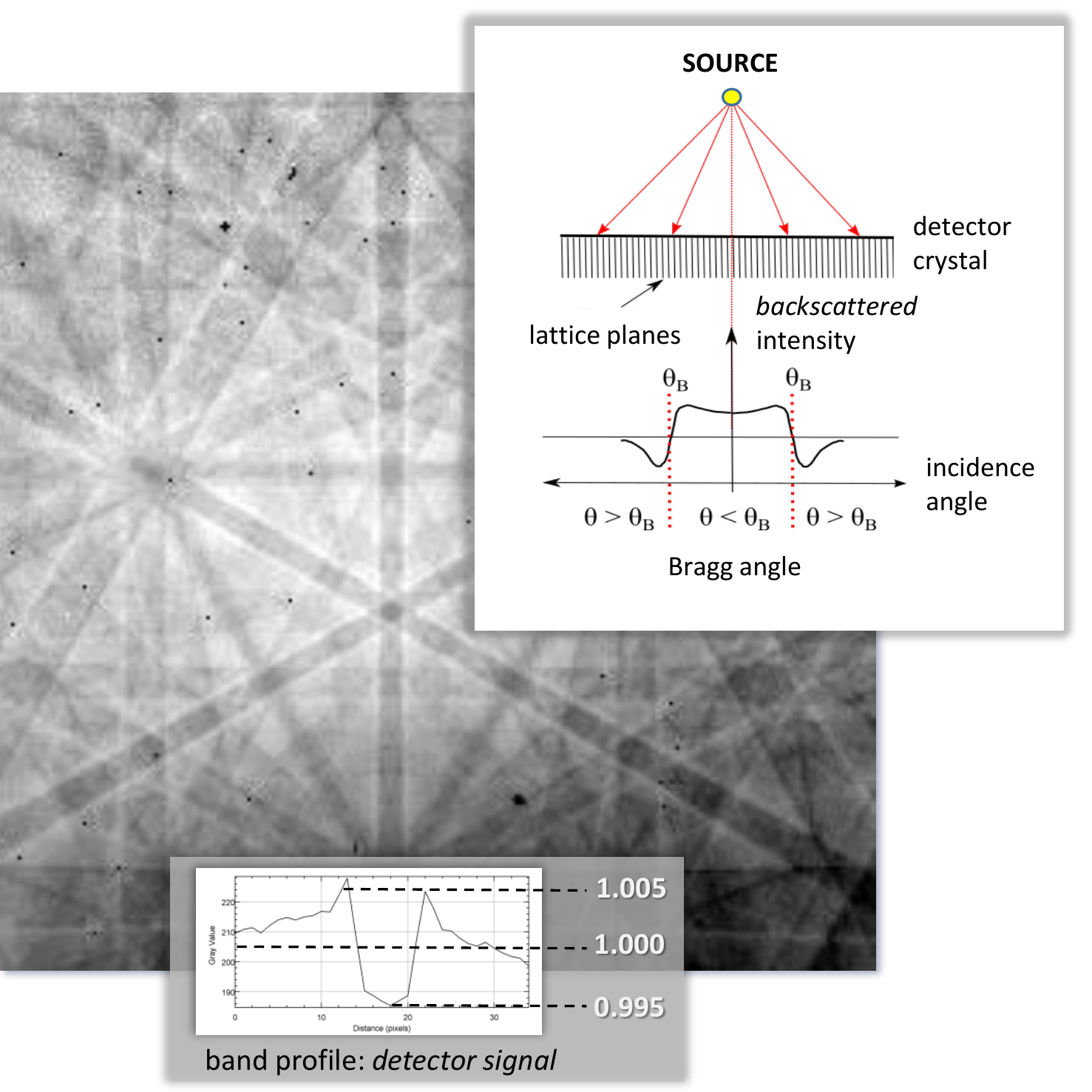}
	\caption{Measured signal on the Timepix detector.
		Pixel(angle)-dependent electron absorption measured on the TimePix detector. Electron channeling effect of electron waves incident on a single-crystalline detector.}
	\label{fig:ddp}
\end{figure}

Summarizing our main idea, the use of an area detector with internal crystalline pixel structure makes it possible to register additional information on the direction of the incoming radiation and to fix the spatial position of a source relative to a detector plane. 
We now demonstrate a practical implementation example of the diffractive ranging method we have discussed above.
Our example is placed in the context of microstructural analysis methods in the scanning electron microscope (SEM). 
Specifically, using the method of electron backscatter diffraction (EBSD), one can obtain information on the crystalline structure and the orientation of the material which is sampled by the electron beam in the SEM.
Backscattered electrons in the SEM form so-called Kikuchi patterns which are measured in a gnomonic projection on a planar screen placed near the sample.  
As an accurate knowledge of the projection center is necessary to calibrate the angular coordinates of the Kikuchi pattern on the detector screen, a key problem in EBSD is the determination of the position of the electron beam spot relative to the detector \cite{biggin1977jac,schwartzEBSD,carpenter2007jm,nolze2007um,engler2009texture,britton2010um,maurice2011um,mingard2011um,basinger2011mm,alkorta2013um,mingard2014iop,chen2015mm,britton2016mc}.

We have previously used a digital hybrid pixel detector, Timepix \cite{llopart2007nima,medipix} in a SEM to obtain Kikuchi patterns from crystalline samples \cite{vespucci2015prb}. 
Detailed investigations of the measured data revealed that the detector response exhibits an underlying diffraction pattern even in the total absence of diffraction effects from the sample, see Fig. \ref{fig:ddp}. 
Strikingly, the observed patterns have a negative intensity distribution relative to what is usually observed for backscattered electrons from a crystal.
As we will show by comparison to simulations, these patterns can be interpreted as electron channeling patterns \cite{joy1982jap} which are formed not by the sample but in the Timepix \textit{detector} crystal instead.

The basic physical mechanism is as follows:
the electrons emitted from the source point (given by the spot where the electron beam from the SEM hits the sample) travel towards the detector plane (made from a Si wafer that is cut in the (111) orientation). 
The electrons impinge on each pixel of the detector from a specific angular direction, e.g. given by the angles in a polar coordinate system with respect to the surface normal and the surface plane.
Due to multiple electron reflection at the lattice planes of silicon, the backscattering probability and penetration depth of the incident electrons into the silicon detector crystal is changed near the Bragg angle due to the preferential excitation of Bloch waves that are localized on lattice planes or between them  \cite{joy1982jap}.
When the number of backscattered electrons is changing, correspondingly the excitation of electron-hole pairs in the silicon pixel element (the measured signal) is varying as a function of incidence angle, see the inset of Fig.\,\ref{fig:ddp}. 
Because less electrons penetrate into the crystal when there is a large backscattered signal, the observed "detector diffraction pattern" (DDP) is inversely proportional to the backscattered intensity.
The  electron channeling effect thus provides a one-to-one relationship between pixel position on the detector and the direction towards the source point as was discussed for the triangulation principle in the introduction.
The specific geometric arrangement of the DDP features on the detector pixels depends  on the position $(x_{\mathrm{P}},y_{\mathrm{P}},z_{\mathrm{P}})$ of the
source point.
A calibrated DDP can thus provide the source point position relative to the screen, as the measured features are fixed by angular relationships in the Si detector crystal, as we will demonstrate below.

\begin{figure}[bt!]   %
	
	\centering
	
	\subfigure[]{
		\includegraphics[width=4.0cm]{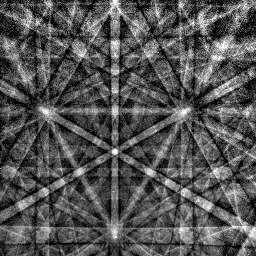}
	}
	\subfigure[]{
		\includegraphics[width=4.0cm]{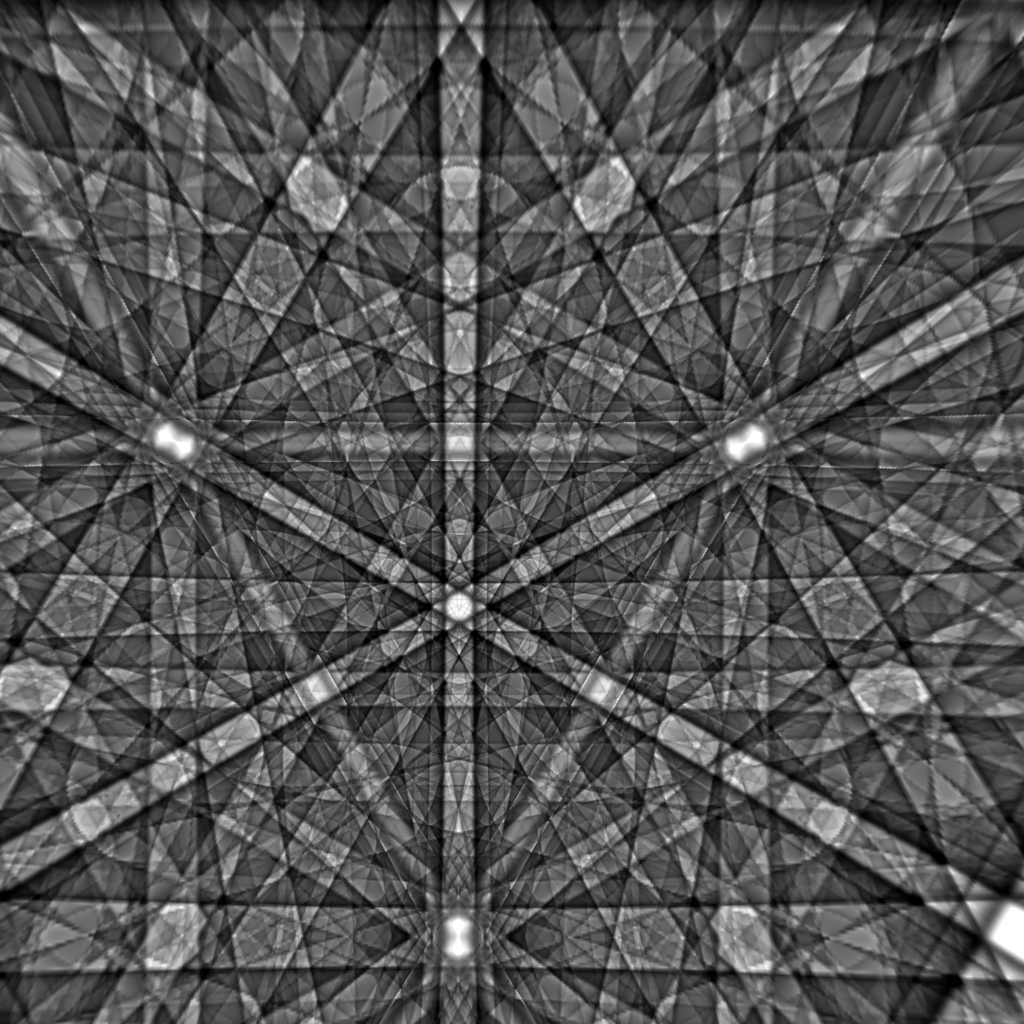}
	}
	\caption{Determination of the source point coordinates from an inverted detector diffraction pattern. The measured pattern for 12 keV beam energy (a) is compared to a dynamical diffraction simulation in (b).
		The best-fit coordinates of the electron source were   
		$(x_P,y_P,z_P)=(6313 \mu m,\; 5753 \mu m,\; 6416 \mu m)$ in the coordinate system of Fig.\,\ref{fig:principle}. The cross-correlation coefficient is $r=0.71$, and the fixed
		detector crystal orientation is ($\phi_1=179.95^\circ$, $\Phi=54.53^\circ$, $\phi_2=45.15^\circ$).}
	\label{fig:calibration}
\end{figure}

In order to determine the 3D position of the electron source with our method, we need to know the exact orientation of the silicon crystal structure comprising the detector device.
The crystal orientation of the sensor material is fixed for the lifetime of the individual detector and is determined by manufacturing variations when cutting the Si crystal into the shape needed for the device. 
Mathematically, the orientation of the Si crystal structure relative to the edges of the detector can be described by Euler angles $(\phi_1, \Phi, \phi_2)$ that describe a rotation sequence around moving ZXZ-axes in the Bunge convention \cite{morawiecOR}.
Ideally, the detector crystal orientation should be determined by an independent method. 
The determination of the detector crystal orientation, however, is inherently limited by the precision of the same type of methods that are also used determine local crystal orientation in an actual sample. 
We have chosen here to estimate the fixed detector orientation as the mean orientation determined from a series of measurements which consist of moving the electron beam in a regular two-dimensional grid over the surface of a sample that shows no backscattering diffraction  pattern.
The calibration procedure for each measured pattern involves a quantitative comparison of the measured DDP with theoretical simulations depending on the source point position and the orientation of the silicon crystal structure with respect to the detector surface plane, see Fig. \ref{fig:calibration} for an example.
%For the nominal 12keV pattern seen in Fig.\,\ref{fig:calibration}, we obtain a best-fit using simulations for 11.3keV and the optimization procedure gives the 3D parameters of the electron source at \textit{fixed} detector orientation as:
%DynamicS values: DD 0,4556731 %PCX  0,4483592 % PCY  0,5913964
%$(x_P,y_P,z_P)=(6313 \mu m, 5753 \mu m, 6416 \mu m)$ in the coordinate system of Fig.\,\ref{fig:principle}.
In a 10x10 map with approx. 10 $\mu$m step size on the sample, the best fit orientation was determined by the optimization of the normalized cross-correlation coefficient $r$ \cite{gonzalez2002} relative to simulated Kikuchi data for Si. For the dynamical electron diffraction simulations \cite{winkelmann2007um} and the best-fit optimizations, we applied the software ESPRIT DynamicS (Bruker Nano, Berlin). %\cite{ESPRITDynamicS}. 

In a first run, both the local orientation and the projection center position were the parameters left to vary in the optimization procedure.
Since we know that the detector crystal is from commercial quality Si wafers, we can  assume a fixed, despite unknown, orientation for the detector. 
Using the open MTEX software \cite{bachmann2010ssp}, this fixed orientation  was approximated as the mean orientation from all the measurements in the map and resulted in Euler angles of  ($\phi_1=179.95^\circ$, $\Phi=54.53^\circ$, $\phi_2=45.15^\circ$). 
This corresponds to a misorientation angle of 0.24$^\circ$ away from an ideal orientation with a \hkl(111) detector surface normal and \hkl[-110] parallel to the horizontal edge of the detector. The size of the deviation is compatible with the overall manufacturing uncertainties.
In order to estimate the orientation variation that is statistically induced in the Si detector crystal simply by the optimization procedure, we calculated an average misorientation of 0.04$^\circ$ with respect to the mean orientation. 
This value shows the precision of the orientation determination due to statistical correlation effects when both the orientation and the projection center is left to vary in the optimization procedure.

In a second optimization step, we then fixed the detector orientation at the  mean orientation from the first run, in order to obtain the final best-fit values for the projection center.
As the beam movement on the sample was in a regular x-y-grid, we can also extract from the measurements an estimation of the repeatability of the projection center determination. 
This is because the projection center $x_P$-coordinate should should also remain constant along a column of the map and the $y_P$ and $z_P$ coordinates stay constant along a row. 
Analysing the mean values in rows and columns of the measured map, we estimate the precision from statistical standard deviations of
$\sigma_{x_P}=2.0\; \mu m$,  $\sigma_{y_P}=1.7\; \mu m$,  $\sigma_{z_P}=1.7\; \mu m$. 

% crystal cut:
% MTEX: MISORIENTATION DETERMINATION
% cs = crystalSymmetry('cubic');
% ss = specimenSymmetry('orthorhombic');
% o_ideal = orientation('Euler',60*degree,54.74356*degree,45*degree,cs,ss)
% best fit orientation of timepix
%o_meas = orientation('Euler',179.946*degree,54.526*degree,45.154*degree,cs,ss)
% angle(o_meas,o_ideal) / degree
% ans = 0.2400

Given the error of $2.0\;\mu m$ estimated above for the projection center coordinates, we can estimate a resulting \textit{absolute} angular error of $\pm0.02^\circ$ for angles near 45$^\circ$ (which in the geometry used here, will be found near the outer boundaries of the detector area).
Applied to strain measurements by EBSD, this estimated angular precision would limit, for example, the \textit{absolute} detection sensitivity for tetragonal changes seen by the change of the 45$^\circ$ angle between a cubic [110] face-diagonal direction and a [001] surface normal direction.
An \textit{absolute} error of $\pm0.02^\circ$ in this angle would correspondingly limit the reliability of the resulting $c/a$-ratio of the unit cell to values of the order of $7\times 10^{-4}$ away from the ideal cubic  $c/a=1.0$.
These \textit{absolute} values are encouraging, when we consider that the conventional strain determination is carried out using pattern resolutions in the order of 1000x1000 pixels with a \textit{relative} strain sensitivity of $10^{-4}$ (i.e. an unknown but fixed projection center and only angular \textit{changes} to be determined).
For an extensive discussion of the strengths and limits of various other existing projection center calibration methods see e.g. \cite{biggin1977jac,britton2010um,maurice2011um,mingard2011um} and references therein.

\begin{figure}   
	\centering
	\subfigure[]{
		\includegraphics[width=4.0cm]{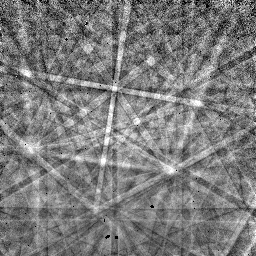}
	}
	\subfigure[]{
		\includegraphics[width=4.0cm]{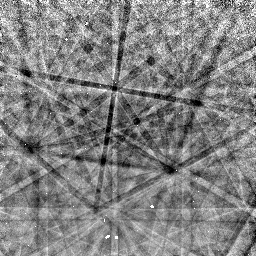}
	}
	\\
	%\vspace{0.1cm}
	\subfigure[]{
		\includegraphics[width=4.0cm]{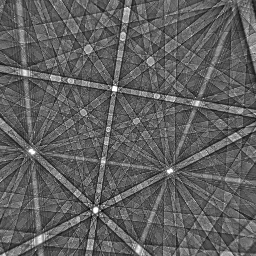}
	}
	\subfigure[]{
		\includegraphics[width=4.0cm]{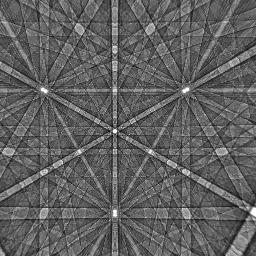}
	}
	\caption{
		Backscattered electron diffraction measurement at 25\,keV obtained from a film of 10nm HfO$_2$ on Si(001) using a Timepix detector.
		(a) Measurement containing simultaneously a visible detector diffraction pattern (dark bands) together with a backscatter Kikuchi pattern (light bands) of the silicon sample.
		(b) Negative of the left pattern \\
		(c) Best-fit simulation of the pattern structure corresponding to the sample orientation of  ($\phi_1=179.95^\circ$, $\Phi=19.93^\circ$, $\phi_2=215.59^\circ$), using the projection center determined in the right panel. 
		(d) Best-fit simulation of the pattern structure corresponding to the detector crystal. This gives the projection point at $(x_P,y_P,z_P)=(6305\;\mu m, 6888 \;\mu m, 6388\; \mu m)$.
	}
	\label{fig:mix}
\end{figure}
Finally, we demonstrate how the watermark-like intensity distribution underlying all the measured data can be used to calibrate an experimental Kikuchi pattern without using any other information other than the pattern itself and the instrumentally fixed detector crystal orientation.
To this end, in Fig.\,\ref{fig:mix}, we present a Kikuchi pattern measurement at 25\,keV using a Si sample covered by 10nm of nanocrystalline HfO$_2$, which for Kikuchi pattern formation can be considered as amorphous. 
The detector crystal orientation was assumed at the values of (179.95$^\circ$, 54.53$^\circ$, 45.15$^\circ$) determined above.
The upper part of Fig.\,\ref{fig:mix} shows the measured pattern (a) and an inverted copy (b) of the same pattern.
The lower part of Fig.\,\ref{fig:mix} shows on the right side (d) the best-fit simulation for the pattern center position from the negative of the total experimental pattern. 
The projection center was determined with a best fit $r$-value of 0.38 at
% (x,y,z) = (0.4477819*14080 ; ()1.0-0.5107961)*14080, 0.4537065*14080) $\mu$m.  
$(x_P,y_P,z_P)=(6305\;\mu m,\; 6888\; \mu m,\; 6388\; \mu m)$.
This corresponds to viewing angles on the detector screen of 95.2$^\circ$ horizontally and 95.5$^\circ$ vertically. 
It is instructive to observe that the cross-correlation approach is reliably detecting the local minimum of $r$ when the simulated pattern registers with that specific part of the pattern structure which is generated only by the detector diffraction.
Finally, we obtained the orientation of the measured sample region by fitting the original measurement in  \mbox{Fig.\,\ref{fig:mix}(a)}, assuming a fixed projection center determined in the previous step from the  inverted pattern in \mbox{Fig.\,\ref{fig:mix}(b)}.
The result is shown in \mbox{Fig.\,\ref{fig:mix}(c)} and corresponds to an orientation of  ($\phi_1=179.95^\circ$, $\Phi=19.93^\circ$, $\phi_2=215.59^\circ$).
The orientation was determined with a best fit $r$-value of $r=0.43$, again showing a selective minimum.
In the optimization procedure using two different patterns mixed in one image, it is actually beneficial that one of the pattern is a negative since this should tend to stabilize the optimization procedure that looks for a maximum of the cross-correlation coefficient, in contrast to a minimum that would be reached for the negative pattern.
In future applications it could be envisaged to combine both optimizations in a simultaneous fit procedure.
In this example experiment, the relative mixture of sample and detector diffraction could be tuned to about 50\% each by adjusting the energy of the electron beam and the thickness of the covering  HfO$_2$ film.
In a conventional experiment involving high-quality crystalline surfaces, the DDP contribution is of the order of parts of a percent, see the inset in Fig. \ref{fig:ddp}.
However, as the detector diffraction contribution is in principle known, the extraction of the DDP "watermark" pattern from the measured Kikuchi pattern should be possible by image processing techniques like template matching or similar approaches \cite{gonzalez2002}. 
Also, one can imagine to produce regular arrays of amorphous reference marks on the sample surface for calibration measurements.

The mode of measurement presented here should also be applicable to other wave sources, given that the source size is sufficiently small compared to the solid angle covered by the detector.
For an electron beam in the SEM, the source size is in the order of 0.1 $\mu m$ for EBSD \cite{zaefferer2007um}.
At distances near 5000 $\mu m$, this corresponds to an angular range of about 0.001 degrees ($2 \times 10^{-5}$\,rad).
As the width of the detector Kikuchi band features is on the order of  several pixels of 55 $\mu m$ dimension, in our case we can still neglect the influence of the source size which will otherwise lead to a blurring of the diffraction features.

%\paragraph{Summary}

In summary, we have discussed a principle of diffractive triangulation of localised radiative sources using crystalline two-dimensional detectors.
As an example, we have demonstrated the application of this principle for the determination of the position of a source of electrons which are backscattered from the surface of a sample in a scanning electron microscope.
For a rigorous crystallographic sample analysis with the highest angular resolution, the precise and accurate knowledge of the incident beam position relative to the planar detector is a necessary prerequisite.
Our approach provides an initial step towards a more accurate determination of the projection center of Kikuchi and other diffraction patterns, which will carry an inherent watermark of the projection center when measured with crystalline active-pixel detectors like Timepix or similar devices \cite{wilkinson2013prl}. 

\appendix{Acknowledgements:}
We would like to thank Joseph Roberts, School of Engineering, University of Liverpool, for providing us with the HfO$_2$ sample and we would like to thank NPL for partial funding of S.V.'s Ph.D. studentship. This work was carried
out with the support of EPSRC Grant No. EP/J015792/1.
 
%\bibliography{ddpbib}
%\bibliography{library}


\end{document}